%% file: main.tex
\documentclass[10pt,conference,letter]{IEEEtran}
\usepackage{graphicx}
\usepackage{epstopdf}
\usepackage{float,url}
\usepackage[caption=false,font=footnotesize]{subfig}
\usepackage{amsmath,epsfig,psfrag,amssymb} 
\usepackage{amsfonts,graphics,graphicx}
\usepackage{algorithm}
\usepackage{algorithmicx,algpseudocode} 
\usepackage{textcomp}
\usepackage{xcolor}

\usepackage{balance,color}

\newtheorem{insight}{Insight}
\newcommand{\ignore}[1]{}

\AtBeginDocument{%
  \providecommand\BibTeX{{%
    \normalfont B\kern-0.5em{\scshape i\kern-0.25em b}\kern-0.8em\TeX}}}

\usepackage{tabularx}
\usepackage{cite}
\begin{document}
\title{Characterizing the Landscape of \\
COVID-19 Themed Cyberattacks and Defenses}

\ignore{
\author{\IEEEauthorblockN{Mir Mehedi Ahsan Pritom}
\IEEEauthorblockA{Department of Computer Science\\
University of Texas at San Antonio\\
Email: mirmehedi.pritom@utsa.edu}
\and
\IEEEauthorblockN{Shouhuai Xu}
\IEEEauthorblockA{Department of Computer Science\\
University of Texas at San Antonio\\
Email: shouhuai.xu@utsa.edu}
\and 
\IEEEauthorblockN{Min Xu}
\IEEEauthorblockA{Mastercard \\
Email: min.xu@mastercard.com}}
}

\author{\IEEEauthorblockN{Mir Mehedi Ahsan Pritom\IEEEauthorrefmark{1},
Kristin M. Schweitzer\IEEEauthorrefmark{2},
Raymond M. Bateman\IEEEauthorrefmark{2},
Min Xu\IEEEauthorrefmark{3} and
Shouhuai Xu\IEEEauthorrefmark{1}}
\IEEEauthorblockA{\IEEEauthorrefmark{1}Department of Computer Science, University of Texas at San Antonio}
\IEEEauthorblockA{\IEEEauthorrefmark{2}U.S. Army Research Laboratory South - Cyber}
\IEEEauthorblockA{\IEEEauthorrefmark{3}Mastercard}
}

\maketitle

\begin{abstract}
COVID-19 (Coronavirus) hit the global society and economy with a big surprise. In particular, work-from-home has become a new norm for employees. Despite the fact that COVID-19 can equally attack innocent people and cyber criminals, it is ironic to see surges in cyberattacks leveraging COVID-19 as a theme, 
dubbed {\em COVID-19 themed cyberattacks} or {\em COVID-19 attacks} for short, which represent a new phenomenon that has yet to be systematically understood. In this paper, we make a first step towards fully characterizing the landscape of these attacks, including their sophistication via the Cyber Kill Chain model. We also explore the solution space of defenses against these attacks.
\end{abstract}
\begin{IEEEkeywords}
COVID-19 Cyberattacks, Malicious Websites, Malicious Emails, Malicious Mobil Apps, Malicious Messaging, Misinformation, Cyber Kill Chain, Defense
\end{IEEEkeywords}

\section{Introduction}

\input{introduction.tex}

\section{Characterizing COVID-19 Attacks}
\label{sec:characterization}

We characterize 5 classes of COVID-19 attacks: malicious websites, malicious emails, malicious mobile apps, malicious messaging, and misinformation. For this purpose, we collect existing news reports and blogs on relevant cyberattacks, manually verify them, and propose mapping them to the Lockheed Martin's Cyber Kill Chain \cite{lockheedMartin_killchain}, which is a model consisting of the following 7 stages.
(i) {\em Reconnaissance}, which corresponds to pre-attack plannings, finding vulnerabilities, collecting possible victims, and setting attack goals.  
(ii) {\em Weaponization}, which corresponds to setting up attack propagation mediums, injecting malicious contents into the mediums, and setting up traps to fool the identified victims.
(iii) {\em Delivery}, which corresponds to the attacker's penetration into a victim's system through some entry point.
(iv) {\em Exploitation}, which corresponds to the wage of actual attacks against a victim's system.
(v) {\em Installation}, which corresponds to the installation of malicious payloads on a victim's system.
(vi) {\em Command-and-Control} (C2), which corresponds to the attacker's use of remote access to the victims' systems.
(vii) {\em Objectives}, which corresponds to the accomplishment of the attacker's pre-determined goal.

\subsection{COVID-19 Themed Malicious Websites}

Attackers have abused websites to wage COVID-19 attacks to steal login credentials,
sell fake medications related to COVID-19, and
inject malicious payloads into these themed websites to distribute malware 
\cite{ten_deadly_covid19_security_threats, CyberCrime_pandemic2020,covid_scam_website_email}. 
We map these attacks to the Cyber Kill Chain model as follows. 

(i) {\em Reconnaissance:} An attacker selects target audience, chooses a COVID-19 related target theme, searches for cheap and unregulated domain registration and web hosting services, and sets attack goals. 
(ii) {\em Weaponization:} An attacker registers new websites with COVID-19 related names.
For example, an attacker may register websites with typo-squatting names to mimic legitimate websites related to COVID-19 (e.g., CDC, WHO, FDA)
\cite{typosquatting_website};
an attacker may register websites to imitate legit Virtual Private Network (VPN) software or remote communication software; an attacker may register domains to offer fake legal services related to  COVID-19; 
an attacker may change an existing phishing website to accommodate COVID-19 themes; 
an attacker may register fake media streaming domains;
and an attacker may register fake donation websites.
(iii) {\em Delivery:} An attacker hosts COVID-19 themed malicious websites mentioned above.
(iv) {\em Exploitation:} Victims visit malicious websites, and then trust the fake forms or download malicious payloads to their devices. 
(v) {\em Installation:} 
A victim may provide sensitive information to a malicious website or intentionally/unintentionally install malware. 
(vi) {\em C2:} An attacker remotely controls victims' infected computers, for example instructing its agents (e.g., malicious websites, downloaded malware) to send the stolen data/credentials to the attacker.
(vii) {\em Objectives:} An attacker gets sensitive credentials,
encrypts a victim's computer,
or gets ransom payment.

\subsection{COVID-19 Themed Malicious Emails}
Attackers have abused emails to wage COVID-19 attacks to send phishing, spamming, scamming, malicious attachments, and malicious websites \cite{email_lures_riskiq}. We map these attacks to the Cyber Kill Chain model as follows. 

(i) {\em Reconnaissance:} An attacker selects target audience, generates and profiles email lists, selects a target topic for COVID-19 themed lures, and sets an attack goal. (ii) {\em Weaponization:} An attacker creates fake typo-squatting email addresses imitating legitimate entities (e.g., CEO, Netflix support team, medical doctors), writes malicious emails with legitimate logo (e.g., WHO, hospital logo) and authority names, writes emails with COVID-19 related information and offers, writes emails with malicious attachments
\cite{malware_ppt_RAT_brazil,fake_VPN_AZORult_malware},
writes fake COVID-19 donation scam emails 
\cite{fake365_for_phishing_CB}, writes emails with fake financial relief payments  \cite{covid_scam_website_email}, writes emails with blackmailing schemes (e.g., threatening languages) 
\cite{covid_email_scam_threats}, writes emails to 
lure victims to provide personal information or pay fees for false unemployment training and certification \cite{covid_job_scams}.   
(iii) {\em Delivery:} An attacker sends the aforementioned emails to the target audience.
(iv) {\em Exploitation:} A victim trusts an email received from an attacker,  
clicks its malicious links, opens its attachments, or downloads its malicious contents. 
(v) {\em Installation:} A victim replies to the attacker with sensitive personal information or 
installs malicious content on its computer either intentionally or unintentionally.
(vi) {\em C2:} An attacker establishes connections with victim's devices through C2 channels, for example, to instruct the compromised computers to send back sensitive data. (vii) {\em Objectives:} An attacker encrypts a victim's computer, receives ransom payment, or receives sensitive information.

\subsection{COVID-19 Themed Malicious Mobile Apps}

Attackers have abused mobile apps to wage COVID-19 attacks to distribute malware and steal information from the victims \cite{COVID_malicious_app_paper}. Google and Apple have taken steps during this pandemic to reject publishing of COVID-19 related mobile apps from unauthorized entities \cite{tech_fight_malicious_apps}. Despite these efforts to secure reputed app stores, malicious apps could still get published and remain undetected as many third party app stores do not have proper reviewing and regulation for publishing apps. Reports showing third-party app stores are eight times more likely to contain malicious apps than than Google Play store \cite{war_on_mal_apps}.  
We map the attack of COVID-19 themed malicious mobile apps to the Cyber Kill Chain model as follows. 

(i) {\em Reconnaissance:} An attacker selects target audience (e.g., based on geographical region), selects a COVID-19 themed topic/service (e.g., tracing, tracking, maps, VPN, remote meeting, COVID-19 guidelines, COVID-19 test information), finds and selects profitable unregulated app stores, and sets attack goals. 
(ii) {\em Weaponization:} An attacker creates fake mobile apps with typo-squatted app names and legitimate logos to imitate authentic apps, repackages existing COVID-19 themed legitimate apps with malware or ransomware (e.g., banking Trojan, spyware) to trick users \cite{COVID_malicious_app_paper}. (iii) {\em Delivery:} An attacker uploads malicious apps into the unregulated app stores or code repositories, and advertises these mobile apps through websites pop-ups. 
(iv) {\em Exploitation:} A victim trusts an malicious app and downloads the app.
(v) {\em Installation:} A victim installs the downloaded malicious app on an mobile device.
(vi) {\em C2:} An attacker remotely controls victims' compromised mobile devices to 
send sensitive user data to the C2 server. 
(vii) {\em Objectives:} An attacker encrypts a victim's mobile device, gets a ransom payment, or steals a victim's private information (e.g., login credentials, crypto wallet passwords), breaches user privacy (e.g., location).

\subsection{COVID-19 Themed Malicious Messaging}
Attackers have abused messaging services to wage COVID-19 attacks (e.g., phishing, malware, spamming, and scamming) \cite{collier2020implications_policy}. COVID-19 has increased the usage of mobile devices which create more incentives for attackers. These attacks are similar to malicious email attacks, but are unique in that messaging can offer more emotional and persuasive live chats. We map them to the Cyber Kill Chain model as follows. 

(i) {\em Reconnaissance:} An attacker selects target audience (e.g., based on demography, geography, severity of COVID-19 infections), collects phone and social media contacts, selects target platform (e.g., Facebook, WhatsApp, Twitter), chooses a COVID-19 themed topic (e.g., fake cures, products, services), and sets attack goals. 
(ii) {\em Weaponization:} An attacker writes persuasive and emotional messages (e.g., asking for COVID-19 donations) to trick victims, creates fake social media profiles, and creates social media groups to lure target audience.
(iii) {\em Delivery:} An attacker sends malicious messages, website links, and attachments through messaging 
to targeted victims, sends scams mentioning fines for leaving home during stay-at-home orders \cite{collier2020implications_policy}, sends fraud messages with free subscription lures for media streaming services 
\cite{ten_deadly_covid19_security_threats}, sends messages to sell low-quality supplies (e.g., masks, gloves, fake cures, and illegal chemical materials) \cite{fake_covid19_product_scam_flood_socialmedia}, sends COVID-19 related lucrative offers (e.g., giveaways, loans, lawyer help, food stamps, stimulus check updates, news guidelines),
and sends crafted misinformation messages with fake claims and made up evidence.
(iv) {\em Exploitation:} A victim trusts a received message and falls victim to it by clicking its malicious links, downloading its malicious contents, and forwarding it to other users.
(v) {\em Installation:} A victim intentionally or unintentionally installs the malicious payload on an messaging device (e.g. Android mobile phone). 
(vi) {\em C2:} An attacker establishes channels (e.g., reply messages, servers connected to a phishing webpage) to remotely control the compromised messaging devices, for example, to receive victims' sensitive information.
(vii) {\em Objectives:} An attacker 
gets victims' sensitive information or
makes lateral movements in victims' networks.

\subsection{COVID-19 Themed Misinformation}

Attackers have waged COVID-19 attacks to spread misinformation, which includes false or inaccurate information (e.g., hoaxes, rumors, or propaganda \cite{sotryful_misinfo}). Examples include: ``COVID-19 is invented in a Chinese lab \cite{wuhanlab_covid19_Vox}"; ``5G is spreading COVID-19 \cite{5g_covid19_verge}",  ``Black are immune to COVID-19 \cite{black_racial_immunity_covid}", ``$X$ can cure COVID-19'' where $X$ can be a drug 
or food items (i.e., Ginger) \cite{WHO_myth_busters}, 
or ``Wearing a mask causes you to inhale too much carbon dioxide, which can make you sick" or ``Wearing a mask can result in getting pneumonia" \cite{debunking_mask_misinformation}. Social media and messaging platforms further increase the impact of such misinformation. The term {\em Infodemic} has even been coined because of this \cite{WHO_met_with_tech}.
We map the COVID-19 themed misinformation attack to the Cyber Kill Chain as follows. 

(i) {\em Reconnaissance:} An attacker analyzes the characteristics of targeted audience (e.g., ethnicity, demography or nationality), identifies vulnerable divisions in society, selects themed topics, and sets attack goals.   
(ii) {\em Weaponization:} An attacker writes fake COVID-19 themed statements and mix them with false evidence and out-of-context truths, creates fake groups in social networking platforms, creates themed memes,
creates bots in social media (e.g., Twitter) to propagate misinformation, and infiltrates into social media groups containing targeted ethnic audience.  
(iii) {\em Delivery:} An attacker posts and shares COVID-19 related misinformation (e.g., narratives, memes, images, and hashtags through social media groups and messaging apps) and publishes fake news on paid online news/tabloids, and/or keeps posting to a larger audience with bots to amplify the impact. 
(iv) {\em Exploitation:} A victims (e.g., social media user) reads and forwards misinformation messages.
(v) {\em Installation:} A victims gets to believe the misinformation which goes viral.  
(vi) {\em C2:} An attacker may generate fake real-life incidents/experience posts on social media related to COVID-19. (vii) {\em Objectives:} An attacker succeeds when bringing more division, mistrust, health crisis, and chaos in society, and possibly earns money from the crisis.

\subsection{Systematizing COVID-19 Themed Cyberattacks}

We systematize COVID-19 attacks by mapping them to their attack techniques and attack goals, and by contrasting their Cyber Kill Chain models. 

\subsubsection{Mapping Attacks, Techniques and Goals}

\begin{figure*}[!htbp]
\centering
\includegraphics[width=0.99\linewidth]{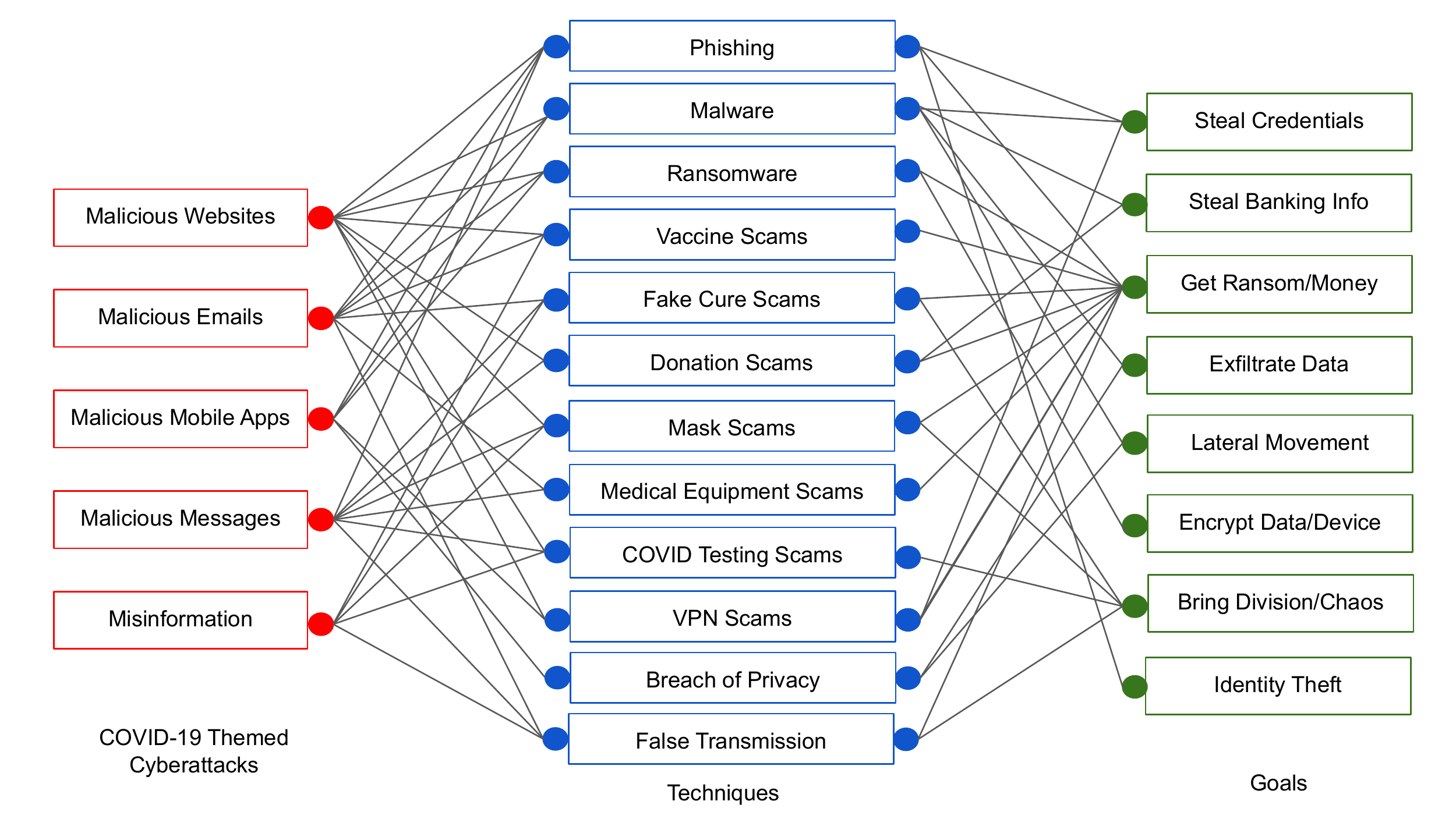}
\caption{Systematizing COVID-19 attacks (red), attack techniques (blue), and attack goals (green)} \label{fig:attack-goal-mapping}
\end{figure*}

Figure \ref{fig:attack-goal-mapping} depicts the mapping between the COVID-19 attacks, the attack techniques they use, and their attack goals. We observe that one attack may use multiple attack techniques. For example, a COVID-19 themed malicious website attack may use a range of attack techniques, including phishing, malware, ransomware, vaccine scams, donation scams, masks scams, testing scams, and VPN scams. Moreover, a COVID-19 themed malicious website attack may have multiple goals. On the other hand, one goal can be achieved by using various kinds of attack techniques, which may be waged through multiple classes of attacks. This means that when an attacker attempt to achieve an attack goal, the attacker can choose attacks and attack techniques in a cost-effective, if not optimal, fashion. For example, each attack may incur some cost or risk (e.g., the cost for using phishing via COVID-19 themed malicious websites and COVID-19 themed malicious emails may be different), and may have different success probabilities (e.g., phishing via COVID-19 themed malicious websites may be more or less successful than phishing via COVID-19 themed malicious emails). This would allow an intelligent attacker to wage the cost-effective or event optimal attack. A systematic framework for achieving this type of attacker decision-making is beyond the scope of the present paper.


\ignore{
{\color{orange}We have observed that most of the COVID-19 themed attacks select one of the scamming techniques to wage the cyberattacks such as, vaccine scams, fake cure scams, donation scams, and so on. However, there are also phishing, malware, and ransomware attack  techniques that select one of the above mentioned themes and wage the attack on the victims through social engineering. 

Again, we observe one of the prior goal of the attacker is to get ransom or monetary benefits which is achieved mostly through different scamming attack techniques. But there are other goals and techniques mappings such as, Phishing: steal credentials, steal banking info, and identity theft; Malware: steal credentials, steal bank info, exfiltrate data, lateral movement, and data breach; Ransomware: encrypt data, encrypt device, lock device, and get ransom. The attackers try to achieve their goals by leveraging the most effective and cost efficient techniques and themes. }

}

\begin{insight}
A COVID-19 attack may use multiple attack techniques to achieve multiple attack goals, and an attack goal may be achieved by using multiple attack techniques that can correspond to multiple attacks. This flexibility allows the attacker to choose cost-effective, if not optimal, attacks in order to achieve a certain attack goal.
\end{insight}

\subsubsection{Systematizing Attacks via Their Cyber Kill Chains}

\begin{figure*}[!htbp]
\centering
\includegraphics[width=1\textwidth]{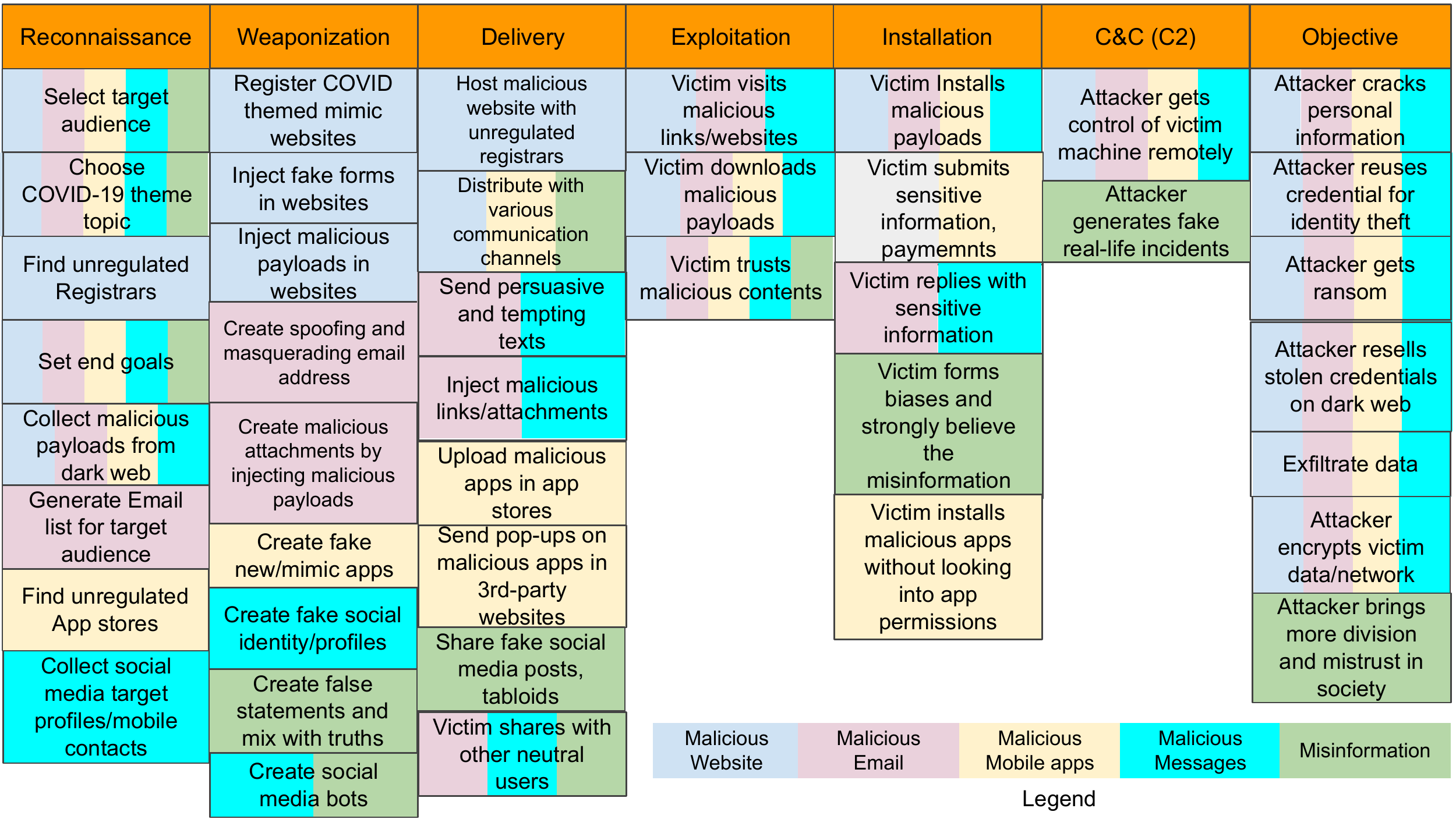}
\caption{Systematizing the Cyber Kill Chains for COVID-19 themed cyberattacks, which are coded in colors (see Legend). 
}
\label{fig:mapping_killchain}
\end{figure*}

Figure \ref{fig:mapping_killchain} depicts the Cyber Kill Chain mappings of the aforementioned 5 classes of COVID-19 attacks, which are represented by different colors. We observe that in each stage of the Cyber Kill Chain, there can be multiple {\em tactics} (e.g., ``select target audience'' and ``choose COVID-19 theme topic'' at the reconnaissance stage). 
We observe that the 5 classes of COVID-19 attacks would use some common tactics at some stages as well as their distinct tactics at other stages. For example, ``select target audience'' at the reconnaissance stage is a tactic that can be used by the 5 classes of attacks, but ``find unregulated app stores'' is a tactic that would be unique to the COVID-19 themed malicious apps attack.  
We also observe that the {\em exploitation} stage almost always leverages victims' mistrust in social engineering, which highlights that human factor remains to be a critical vulnerability in COVID-19 attacks, which reinforces the importance of seeking effective defenses against such attacks \cite{DBLP:journals/corr/abs-2007-04932}. 

\begin{insight}
COVID-19 attacks can be very sophisticated, rather than only opportunistic, which means that effective defense must be designed on a deeper understanding about the attack tactics that can be used in each stage of the attack (i.e., knowing the attacker better).
\end{insight}

\section{Exploring the Defense Space}
\label{sec:defense}
\input{defense.tex}

\section{Conclusion}
\label{sec:conclusion}

We have explored the landscape of COVID-19 themed cyberattacks and defenses. We discussed 5 classes of attacks and mapped them to the Cyber Kill Chain model. 
We explored defense strategies against these attacks.
Although the study is geared towards COVID-19 themed cyberattacks, the exploration and landscape can be adapted to future $X$-themed cyberattacks exploiting future events (e.g., election, natural or man-made disasters). It is also interesting to rigorously model these attack-defense interactions in the Cybersecurity Dynamics framework \cite{XuCybersecurityDynamicsHotSoS2014,XuAgility2019,XuPLoSOne2015,XuTIFSDataBreach2018}.

\smallskip

\noindent{\bf Acknowledgement}. We thank the reviewers for their useful comments. This work was supported in part by ARO Grant \#W911NF-17-1-0566, ARL Grant \#W911NF-17-2-0127, and the NSA OnRamp II program.

\bibliographystyle{IEEEtran}
\bibliography{sample-base,metrics}

\end{document}

%% file: introduction.tex

The COVID-19 (Cronavirus) pandemic has had a huge impact on the global society and economy. It attacks everyone, including both the innocent people and the cyber criminals. Ironically, we have witnessed surges in cyberattacks leveraging COVID-19 as a theme, dubbed {\em COVID-19 themed cyberattacks} or {\em COVID-19 attacks} for short. 
For example, there is a 32X increase in the malware and phishing sites from February 25, 2020 to March 25, 2020 \cite{COVID_attack_phishing_menlo}; 
Google has been
blocking 240 million COVID-19 related spam emails and 18 million phishing and malware emails daily \cite{Google_TAG}; there is a 148\% increase in ransomware attacks in March 2020 over February 2020 \cite{VMware_carbonBlack_report}. 
The situation is further exacerbated by the new norm of work-from-home because employees' home computers or devices are often less protected than their enterprise counterparts.
Indeed, a CheckPoint survey \cite{Checkpoint_survey} shows that 55\% of security professionals are concerned with remote access
and 47\% are concerned with their employees using shadow IT systems from their home. Until now, COVID-19 attacks have mainly targeted the finance, healthcare, government, media streaming, retail business, and COVID-19 research sectors. In response, experts have recommended using multi-factor authentication for critical transactions, virtual private networks for remote access, and regularly patching and updating software as immediate solutions \cite{CBWashingHand20s}.
However, given that COVID-19 attacks are a new phenomenon that is here to stay, it is important to understand them thoroughly to pave a way for effective defense.

\smallskip

\noindent{\bf Our contributions}.
In this paper, we make a first step towards understanding COVID-19 themed cyberattacks. Specifically, we explore five classes of them, namely malicious websites, malicious emails, malicious mobile apps, malicious messaging, and misinformation. In order to characterize these attacks, we map them to the Cyber Kill Chain model \cite{lockheedMartin_killchain}. We show that they can use multiple attack techniques to achieve multiple attack goals. We find that  COVID-19 attackers have been professional rather than opportunistic and have been heavily employing social-engineering cyberattacks. 
We further explore the solution space of defenses against COVID-19 attacks. Since COVID-19 attacks do have their counterparts that are not specific to the COVID-19 incidents, our focus is on exploring the COVID-19 specific aspects. 
To the best of our knowledge, this is the first systematic characterization of COVID-19 attacks and defenses, which can be adapted to cope with any ``$X$-themed cyberattacks'' that may emerge in the future, where $X$ can be any kind of social incidents (e.g., election, natural or man-made disaster).

\smallskip

\noindent{\bf Related work}.
The problem of COVID-19 attacks has started to receive attention from the research community. There are studies on the types of cyberattacks and their trends amid the COVID-19 pandemic 
\cite{ten_deadly_covid19_security_threats,CyberCrime_pandemic2020,collier2020implications_policy}, studies on specific cyberattacks related to COVID-19 (e.g., mobile malware ecosystem \cite{COVID_malicious_app_paper,lallie2020cyber_COVID19Paper}, 
cybercrimes \cite{influence_model_COVID_cybercrime}, fake news related to COVID-19 \cite{CyberCrime_in_time_of_plauge}, and COVID-19 themed cryptocurrency scams \cite{xia2020dont_cryptocurrency}).
When compared with these studies, we aim at systematically characterizing the landscape of the COVID-19 attacks, including their sophistication through the the Kill Chain \cite{lockheedMartin_killchain} and exploring the space of defenses against these attacks.

\smallskip

\noindent{\bf Paper outline}. 
Section \ref{sec:characterization} characterizes COVID-19 attacks. Section \ref{sec:defense} explores the defense solution space. Section \ref{sec:conclusion} concludes the paper.

%% file: defense.tex
The preceding characterization of COVID-19 attacks guides us to explore defense strategies against them, with an emphasis on {\em what-to-leverage} when designing  defense systems. The investigation of these proposed approaches is beyond the scope of the present paper. This is because each approach needs to be investigated separately, with corresponding experiments.

\subsection{COVID-19 Malicious Websites Defense}

We propose four approaches to defending against COVID-19 themed malicious websites.
The first approach is to leverage various website contents pertinent to COVID-19. 
What is unique to content-based detection of COVID-19 themed malicious websites is the COVID-19 related features, such as the presence or absence of keywords in website names (e.g., {\em coronavirus}, {\em COVID-19}, {\em masks}, {\em n95}, and {\em test}).  
The second approach is to leverage website environment, including URLs' information.
For example, 
typo-squatting URLs or mimicking fake websites can be detected by analyzing URLs information and website screenshots. 
The third approach is to leverage websites' age information. Since COVID-19 themed malicious websites would be created after the outbreak of the COVID-19 pandemic, hinting that the lifetime of many such websites would be short.
The fourth approach is to leverage effective training to make users more skeptical about website contents.

\subsection{COVID-19 Malicious Emails Defense}

We propose three approaches to defending against COVID-19 themed malicious emails. The first approach is to filter emails by searching COVID-19 themed keywords in their subject lines and contents. Examples of such keywords include: {\em COVID-19 cures}, {\em COVID-19 guidelines}, and {\em COVID-19 offers}. The second approach is to verify the sender email address to detect email masquerading \cite{email_masquerading}. The third approach is to leverage email content, for example by analyzing their attachments, links and texts.

\subsection{COVID-19 Malicious Mobile Apps Defense}
We propose four approaches to defending against COVID-19 themed malicious apps. The first approach is to leverage computer vision to proactively examine newly published app's logos, especially when they are similar to, if not exactly the same as, the logos of some popular legitimate apps.
The second approach is to analyze the content of apps to detect the malicious ones (e.g., repackaged apps). For this purpose, static analysis, dynamic analysis, and their combinations may be utilized. The third approach is to examine the string edit distance of app names with respect to some popular ones. The fourth approach is to train users to improve their awareness of malicious apps according to some best practices in using mobile apps securely \cite{MobileApp_bestpractice_users}.

\subsection{COVID-19 Malicious Messaging Defense}
We propose three approaches to defending against COVID-19 themed malicious messaging. The first approach is to leverage message content to check if a message contains suspicious content (e.g., the presence of URLs, emoticons, special characters, and COVID-19 themed keywords).
The second approach is to detect persuasive messages waging social engineering cyberattacks. This may be achieved by analyzing texts and leveraging human factors and psychological means \cite{DBLP:journals/corr/abs-2007-04932}. 
The third approach is to train users to improve their awareness of COVID-19 themed malicious messages.

\subsection{COVID-19 Misinformation Defense}
We propose four approaches to defending against COVID-19 themed misinformation attacks. The first approach is to use fact-checking to detect fake news (or social media posts), perhaps using similar news reports from credible sources and AI or machine learning techniques. 
The second approach is to use central repositories to host COVID-19 related information and resources (e.g., Facebook's COVID-19 Information Center).
The third approach is to train and educate users to improve their skills and capabilities in recognizing fake misinformation. 
The fourth approach is to leverage crowdsourcing, namely encouraging or incentivizing users to report COVID-19 suspicious misinformation posts and links.

%% file: main.bbl
\begin{thebibliography}{10}
\providecommand{\url}[1]{#1}
\csname url@samestyle\endcsname
\providecommand{\newblock}{\relax}
\providecommand{\bibinfo}[2]{#2}
\providecommand{\BIBentrySTDinterwordspacing}{\spaceskip=0pt\relax}
\providecommand{\BIBentryALTinterwordstretchfactor}{4}
\providecommand{\BIBentryALTinterwordspacing}{\spaceskip=\fontdimen2\font plus
\BIBentryALTinterwordstretchfactor\fontdimen3\font minus
  \fontdimen4\font\relax}
\providecommand{\BIBforeignlanguage}[2]{{%
\expandafter\ifx\csname l@#1\endcsname\relax
\typeout{** WARNING: IEEEtran.bst: No hyphenation pattern has been}%
\typeout{** loaded for the language `#1'. Using the pattern for}%
\typeout{** the default language instead.}%
\else
\language=\csname l@#1\endcsname
\fi
#2}}
\providecommand{\BIBdecl}{\relax}
\BIBdecl

\bibitem{COVID_attack_phishing_menlo}
Menlo-Security, ``Sophisticated covid-19–based phishing attacks leverage pdf
  attachments and saas to bypass defenses,''
  https://www.menlosecurity.com/blog/sophisticated-covid-19-based-phishing-attacks-leverage-pdf-attachments-and-saas-to-bypass-defenses,
  2020, accessed on 5th June, 2020.

\bibitem{Google_TAG}
S.~Huntley, ``Findings on covid-19 and online security threats,''
  https://www.blog.google/threat-analysis-group/findings-covid-19-and-online-security-threats/,
  2020, accessed on 5th June, 2020.

\bibitem{VMware_carbonBlack_report}
P.~Upatham and J.~Treinen, ``Amid covid-19, global orgs see a 148\% spike in
  ransomware attacks; finance industry heavily targeted,''
  https://www.carbonblack.com/2020/04/15/amid-covid-19-global-orgs-see-a-148-spike-in-ransomware-attacks-finance-industry-heavily-targeted/,
  2020, accessed on 10 June, 2020.

\bibitem{Checkpoint_survey}
CheckPoint, ``A perfect storm: the security challenges of coronavirus threats
  and mass remote working,''
  https://www.blog.google/threat-analysis-group/findings-covid-19-and-online-security-threats/,
  2020, accessed on 5th June, 2020.

\bibitem{CBWashingHand20s}
R.~McElroy, ``What is the cyber security equivalent of washing your hands for
  20 seconds?''
  https://www.enterprisetimes.co.uk/2020/04/15/what-is-the-cyber-security-equivalent-of-washing-your-hands-for-20-seconds/,
  2020, accessed on 31st May, 2020.

\bibitem{lockheedMartin_killchain}
E.~M. Hutchins, M.~J. Cloppert, and R.~M. Amin, ``Intelligence-driven computer
  network defense informed by analysis of adversary campaigns and intrusion
  kill chains,'' \emph{Leading Issues in Information Warfare \& Security
  Research}, vol.~1, p.~80, 2011.

\bibitem{ten_deadly_covid19_security_threats}
\BIBentryALTinterwordspacing
N.~A. Khan, S.~N. Brohi, and N.~Zaman, ``Ten deadly cyber security threats amid
  covid-19 pandemic,'' May 2020. [Online]. Available:
  \url{https://www.techrxiv.org/articles/Ten_Deadly_Cyber_Security_Threats_Amid_COVID-19_Pandemic/12278792/1}
\BIBentrySTDinterwordspacing

\bibitem{CyberCrime_pandemic2020}
M.~V. Fontanilla, ``Cybercrime pandemic,'' \emph{Eubios Journal of Asian and
  International Bioethics}, vol.~30, no.~4, pp. 161--165, 2020.

\bibitem{collier2020implications_policy}
B.~Collier, S.~Horgan, R.~Jones, and L.~A. Shepherd, ``The implications of the
  covid-19 pandemic for cybercrime policing in scotland: a rapid review of the
  evidence and future considerations,'' 2020.

\bibitem{COVID_malicious_app_paper}
R.~He, H.~Wang, P.~Xia, L.-L. Wang, Y.~Li, L.~Wu, Y.~Zhou, X.~Luo, Y.~Guo, and
  G.~Xu, ``Beyond the virus: A first look at coronavirus-themed mobile
  malware,'' \emph{ArXiv}, vol. abs/2005.14619, 2020.

\bibitem{lallie2020cyber_COVID19Paper}
H.~S. Lallie, L.~A. Shepherd, J.~R.~C. Nurse, A.~Erola, G.~Epiphaniou,
  C.~Maple, and X.~Bellekens, ``Cyber security in the age of covid-19: A
  timeline and analysis of cyber-crime and cyber-attacks during the pandemic,''
  2020.

\bibitem{influence_model_COVID_cybercrime}
R.~Naidoo, ``A multi-level influence model of covid-19 themed cybercrime,''
  \emph{European Journal of Information Systems}, pp. 1--16, 2020.

\bibitem{CyberCrime_in_time_of_plauge}
K.~Gradoñ, ``Crime in the time of the plague: Fake news pandemic and the
  challenges to law-enforcement and intelligence community,'' \emph{Society
  Register}, vol.~4, no.~2, pp. 133--148, 2020.

\bibitem{xia2020dont_cryptocurrency}
P.~Xia, H.~Wang, X.~Luo, L.~Wu, Y.~Zhou, G.~Bai, G.~Xu, G.~Huang, and X.~Liu,
  ``Don't fish in troubled waters! characterizing coronavirus-themed
  cryptocurrency scams,'' 2020.

\bibitem{covid_scam_website_email}
T.~Brewster, ``Coronavirus scam alert: Watch out for these risky covid-19
  websites and emails,''
  https://www.forbes.com/sites/thomasbrewster/2020/03/12/coronavirus-scam-alert-watch-out-for-these-risky-covid-19-websites-and-emails/,
  2020, accessed on 11 June, 2020.

\bibitem{typosquatting_website}
PrivSec, ``Typosquatting \& duplication of pharmaceutical domain – possibly
  used for phishing activity,''
  https://gdpr.report/news/2020/05/06/typosquatting-duplication-of-pharmaceutical-domain-possibly-used-for-phishing-activity/,
  2020, accessed on 9 June, 2020.

\bibitem{email_lures_riskiq}
RiskIQ, ``Covid-19 cybercrime update,''
  "https://www.riskiq.com/blog/analyst/covid19-cybercrime-update/", 2020,
  accessed on 3 June, 2020.

\bibitem{malware_ppt_RAT_brazil}
S.~Singh, ``Coronavirus-themed document targets brazilian users,''
  https://www.zscaler.com/blogs/research/coronavirus-themed-document-targets-brazilian-users,
  2020, accessed on 10 June, 2020.

\bibitem{fake_VPN_AZORult_malware}
S.~Gatlan, ``Azorult malware infects victims via fake protonvpn installer,''
  https://www.bleepingcomputer.com/news/security/azorult-malware-infects-victims-via-fake-protonvpn-installer/,
  2020, accessed on 5 June, 2020.

\bibitem{fake365_for_phishing_CB}
Ac, J.~Myers, and E.~Murphy, ``Technical analysis: Hackers leveraging covid-19
  pandemic to launch phishing attacks, fake apps/maps, trojans, backdoors,
  cryptominers, botnets \& ransomware,''
  https://www.carbonblack.com/2020/03/19/technical-analysis-hackers-leveraging-covid-19-pandemic-to-launch-phishing-attacks-trojans-backdoors-cryptominers-botnets-ransomware/,
  2020, accessed on 5th June, 2020.

\bibitem{covid_email_scam_threats}
B.~Small, ``Scam emails demand bitcoin, threaten blackmail,''
  https://www.consumer.ftc.gov/blog/2020/04/scam-emails-demand-bitcoin-threaten-blackmail,
  2020, accessed on 11 June, 2020.

\bibitem{covid_job_scams}
L.~W. Schifferle, ``Looking for work after coronavirus layoffs?''
  https://www.consumer.ftc.gov/blog/2020/04/looking-work-after-coronavirus-layoffs,
  2020, accessed on 11 June, 2020.

\bibitem{tech_fight_malicious_apps}
I.~Sherr, ``Apple, google, amazon block nonofficial coronavirus apps from app
  stores,''
  https://www.cnet.com/news/apple-google-amazon-block-nonofficial-coronavirus-apps-from-app-stores/,
  2020, accessed on 2nd June, 2020.

\bibitem{war_on_mal_apps}
R.~Lemons, ``Lessons from the war on malicious mobile apps,''
  https://www.darkreading.com/mobile/lessons-from-the-war-on-malicious-mobile-apps/d/d-id/1333946,
  2019, accessed on 3rd June, 2020.

\bibitem{fake_covid19_product_scam_flood_socialmedia}
R.~Heilweil, ``Coronavirus scammers are flooding social media with fake cures
  and tests,''
  https://www.vox.com/recode/2020/4/17/21221692/digital-black-market-covid-19-coronavirus-instagram-twitter-ebay,
  2020, accessed on 11 June, 2020.

\bibitem{sotryful_misinfo}
Storyful-Intelligence, ``Misinformation and disinformation,''
  https://storyful.com/thought-leadership/misinformation-and-disinformation/,
  2018, accessed on 5th June, 2020.

\bibitem{wuhanlab_covid19_Vox}
E.~Barclay, ``Why these scientists still doubt the coronavirus leaked from a
  chinese lab,''
  https://www.vox.com/2020/4/23/21226484/wuhan-lab-coronavirus-china, 2020,
  accessed on 5th June, 2020.

\bibitem{5g_covid19_verge}
J.~Vincent, ``Conspiracy theorists say 5g causes novel coronavirus, so now
  they’re harassing and attacking uk telecoms engineers,''
  https://www.theverge.com/2020/6/3/21276912/5g-conspiracy-theories-coronavirus-uk-telecoms-engineers-attacks-abuse,
  2020, accessed on 5th June, 2020.

\bibitem{black_racial_immunity_covid}
J.~Ross, ``Coronavirus outbreak revives dangerous race myths and
  pseudoscience,''
  https://www.nbcnews.com/news/nbcblk/coronavirus-outbreak-revives-dangerous-race-myths-pseudoscience-n1162326,
  2020, accessed on 12 June, 2020.

\bibitem{WHO_myth_busters}
W.~H. Organization, ``Coronavirus disease (covid-19) advice for the public:
  Myth busters,''
  https://www.who.int/emergencies/diseases/novel-coronavirus-2019/advice-for-public/myth-busters,
  2020, accessed on 12 June, 2020.

\bibitem{debunking_mask_misinformation}
Intermountain-Healthcare, ``Debunking common face mask misconceptions,''
  https://intermountainhealthcare.org/blogs/topics/covid-19/2020/07/debunking-common-face-mask-misconceptions/,
  07 2020, accessed on 1st August, 2020.

\bibitem{WHO_met_with_tech}
S.~R. Christina~Farr, ``Facebook, amazon, google and more met with who to
  figure out how to stop coronavirus misinformation,''
  https://www.cnbc.com/2020/02/14/facebook-google-amazon-met-with-who-to-talk-coronavirus-misinformation.html,
  2020, accessed on 4th June, 2020.

\bibitem{DBLP:journals/corr/abs-2007-04932}
R.~M. Rodriguez, E.~Golob, and S.~Xu, ``Human cognition through the lens of
  social engineering cyberattacks,'' \emph{Frontiers in Psychology}, 2020.

\bibitem{email_masquerading}
S.~Baki, R.~Verma, A.~Mukherjee, and O.~Gnawali, ``Scaling and effectiveness of
  email masquerade attacks: Exploiting natural language generation,'' in
  \emph{Proc. ACM AsiaCCS}, 2017, p. 469–482.

\bibitem{MobileApp_bestpractice_users}
Y.~Magali, ``10 security best practices for mobile device owners,''
  https://www.cdillc.com/10-security-best-practices-mobile-device-owners/,
  accessed on 8 August, 2020.

\bibitem{XuCybersecurityDynamicsHotSoS2014}
S.~Xu, ``Cybersecurity dynamics,'' in \emph{Proc. HotSoS'14}, 2014, pp.
  14:1--14:2.

\bibitem{XuAgility2019}
J.~Mireles, E.~Ficke, J.~Cho, P.~Hurley, and S.~Xu, ``Metrics towards measuring
  cyber agility,'' \emph{IEEE T-IFS}, vol.~14, no.~12, pp. 3217--3232, 2019.

\bibitem{XuPLoSOne2015}
Y.~Chen, Z.~Huang, S.~Xu, and Y.~Lai, ``Spatiotemporal patterns and
  predictability of cyberattacks,'' \emph{PLoS One}, vol.~10, no.~5, p.
  e0124472, 05 2015.

\bibitem{XuTIFSDataBreach2018}
M.~Xu, K.~M. Schweitzer, R.~M. Bateman, and S.~Xu, ``Modeling and predicting
  cyber hacking breaches,'' \emph{IEEE T-IFS}, vol.~13, no.~11, pp. 2856--2871,
  2018.

\end{thebibliography}
